\documentstyle[aps]{revtex}
\begin{document}
\title{Thermodynamic identities and particle number fluctuations \\
       in weakly interacting Bose--Einstein condensates}
\author{ 
Fabrizio Illuminati$^{*} \S $,
Patrick Navez$^{*} \dagger $ and
Martin Wilkens$^{*}$}

\address{
$^{*}$ Institut f\"ur Physik, Universit\"at Potsdam, \\
Am Neuen Palais 10; D--14415, Potsdam, Germany. \\
$\S $ Dipartimento di Fisica, Universit\`a di Salerno, and INFM, \\
Unit\`a di Salerno, I--84081 Baronissi (SA), Italy. \\
$\dagger $ Institute of Materials Science, Demokritos N.C.S.R., 
POB 60228, 15310 Athens, Greece.}

\date{May 10, 1999}

\maketitle

\begin{abstract}

We derive exact thermodynamic identities relating 
the average number of condensed atoms and the 
root--mean--square fluctuations determined in different
statistical ensembles for the weakly interacting Bose
gas confined in a box. This is achieved by introducing
the concept of {\it auxiliary partition functions} for 
model Hamiltonians that do conserve the total number 
of particles.
Exploiting such thermodynamic identities, we
provide the first, completely analytical
prediction of the microcanonical particle number
fluctuations in the weakly interacting Bose gas.
Such fluctuations, as a function of the volume
$V$ of the box  are found to behave normally,
at variance with the anomalous scaling behavior
$V^{4/3}$ of the fluctuations in the ideal Bose
gas.  

\end{abstract}

\vspace{0.4cm}

After the experimental achievement of Bose--Einstein condensation
in trapped, cold gases of weakly interacting
neutral Bose atoms \cite{And}--\cite{Dav}, a flow of
experimental data on the statistical properties of the condensate
fraction has become available for theoretical study
\cite{Dalf}. 
Among others, the issue of the fluctuations in the number of 
condensed atoms is of central importance, and it is foreseeable
that it will become experimentally testable in the near future.

Experimentally, the trapped atoms are in a situation of almost
complete isolation with respect to the outer environment surrounding
the trap: the total number of particles is certainly
conserved, while energy exchange with the environment is 
small, in some cases even negligible. 
Therefore, the relevant statistical ensemble to consider 
when attempting to compute the fluctuations should
be the canonical or, even more appropriate,
the microcanonical one.

For the ideal Bose gas, the fluctuations of the condensate fraction
were calculated in 
a series of papers \cite{Gross1}-\cite{Wil}, both
in the microcanonical and the canonical
ensembles by a variety of approximate methods:
both numerical (based on recurrence formulae and on contour
integration) and analytic (asymptotic formulae based on the notion of
the Maxwell Demon ensemble), for various trapping potentials and in
different number of dimensions.

On the other hand, the issue of fluctuations in the weakly 
interacting Bose gas is not as well defined, both because
there is no unambigous way of defining the condensate
fraction and because the approximate dynamics and energy
spectra can be defined in a number of different ways, yielding
different model--dependent predictions.

Of the few existing attempts to address the problem, we should
mention in particular the calculation of the
fluctuations of the interacting Bose gas trapped in a box, 
that was developed analytically in the canonical ensemble 
assuming the
Bogoliubov number nonconserving description
\cite{Gior}. In this framework, it was
predicted an anomalous scaling of the 
fluctuations with the volume $V$ of the box,
$<\delta^{2}N_{0}>_{cn} \sim V^{4/3}$ analogous to that
of the ideal Bose gas. However, this result can
be questioned on the grounds that it has been
derived assuming the condensate to act as a particle
reservoir, and moreover that in the actual 
experimental situations the total number of
particles $N$ is a good quantum number. 

It is therefore sensible to set up model
calculations that avoid these features, and
to compare the results thus obtained.
As a first step, the fluctuations were calculated
analytically in the canonical ensemble
and numerically in the microcanonical ensemble, 
assuming different standard
approximate spectra (respectively, the one obtained in
Hartree--Fock theory, the two--gas model obtained by
reduction of the Bogoliubov approximation, and the spectrum
obtained in lowest order perturbation theory). 
It was shown
that the fluctuations are very sensitive to the details of
the approximation, resulting in large discrepancies between 
the different models \cite{Kazik}. 
On the other hand, they all satisfy
the criterion of conservation of the total number
of particles, while the lowest order perturbative
spectrum includes three-- and four--body contributions
to the dynamics.

In this letter we address the issue of computing analytically
the fluctuations of the condensate fraction in the microcanonical
ensemble for a specific number--conserving model interaction.
By introducing a so--called auxiliary canonical partition
function, defined below, 
we prove by saddle point evaluation and with the help of 
standard thermodynamic Legendre transformations that 
for all model Hamiltonians that do conserve the total
number of particles, the 
microcanonical fluctuations can be expressed in terms of the
canonical moments exactly as in the case of the microcanonical
fluctuations in the ideal Bose gas \cite{Nav}.

On the other hand, the canonical moments are modified in the
presence of the interaction, and so is the final analytic expression
of the microcanonical fluctuations, which we are able to compute.
At small but finite coupling (scattering length)
the fluctuations scale linearly with the volume $V$ of the box,
showing no anomalous behavior. 
In all actual calculations we assume for the 
interacting system the energy spectrum 
obtained in lowest order perturbation theory \cite{Huang}.

The Bose--Einstein condensation manifests itself in the macroscopic
occupation of one of the eigenmodes. In general, the eigenfunctions of
the single particle density matrix depend on the total number of atoms
$N$, on the trapping potential, on the interatomic interactions and on
the statistical ensemble of the system. 
The exceptionally simple, and hence the
best studied, is the case of a system confined in the box with
periodic boundary conditions. In this case the single particle density
matrix is translationally invariant and periodic, so its spectral
decomposition is just the Fourier series, and the index
$j$ is the quantized momentum ${\bf p} = (2 \hbar
\pi/L)(n_x,n_y,n_z)$, where $L$ is the size of the box. 

We are interested in the statistics of a weakly interacting Bose gas
for temperatures below the condensation temperature. In this regime,
the system may be viewed as being composed of two macroscopic
subsystems, a condensate of $N_0\sim O(N)$ particles occupying the
${\bf p}=0$ state, and an excited part of $N_{ex}\equiv\sum_{{\bf
p}\neq0}n_{\bf p}$ particles. For isolated systems, particle number
conservation implies $N_{ex}=N-N_{0}$.

After splitting the system into its condensed and excited parts we
will have to define the approximate dynamics. 
This needs not to be specified as long as we are concerned with
general thermodynamic arguments, but only in the actual calculations. 

One commonly made assumption is to neglect the interatomic interaction
within the excited subsystem. Thus, the states of the excited atoms
are still single particle states in a box, since the condensate is
uniform. Accordingly, the energy of the excited subsystem is given by
\begin{equation}
E_{ex} = \sum_{{\bf p}\neq0}\frac{{\bf p}^{2}}{2m}n_{\bf p}\,.
\end{equation}

To fully define the problem, we need to specify the relation between
the total energy of the system $E$ and $E_{ex}$, subject to the
requirements that the total number of atoms $N$ be a good quantum
number, and that higher order contributions beyond the two--body
ones be included. We will hence adopt 
the spectrum obtained in lowest order perturabtion theory \cite{Huang}
in its simplified form by neglecting the terms proportional to
the product of excited occupation numbers:
\begin{equation}
E(N, N_{ex}) = \alpha (N^2 + 2 NN_{ex} - N_{ex}^2) + E_{ex} \, ,
\end{equation}

\noindent where 
\begin{equation}
\alpha = \frac{2 \pi a \hbar^2}{m V} \, ,
\end{equation}

\noindent and $a$ is the scattering length. 

The above, like the other mentioned approximate spectra, depends
only on the total number of atoms $N$ and on the total number
of excited atoms $N_{ex}$ (or, equivalently, on $N$ and $N_{0}$),
and not on the detailed distribution among excited states. 

To calculate the canonical expressions for mean values and 
fluctuations and to relate them to the microcanonical ones 
in an exact thermodynamic
setting, we may now proceed as follows: 
we define an auxiliary canonical
partition function for exactly $N$ particles, by introducing
the chemical potential $\mu_{ex}$ pertaining to the subsystem
of excited atoms:
\begin{equation}
Z_{aux}(N, \beta , \mu_{ex}) = 
Tr_{N} \left[ \exp{-(\beta H - \mu_{ex}N_{ex})} \right] =
\sum_{N_{ex}} \exp{[-\mu_{ex}N_{ex}]}Z_{ex}(N,N_{ex},\beta ) \, .
\end{equation}

The crucial observation is now that differentiating the 
free energy with respect
to $\mu_{ex}$ and evaluating the derivative at $\mu_{ex} =0$
yields the canonical average $<N_{0}>_{cn}$. Furthermore,
since $N_{ex}$ commutes with the total Hamiltonian, second derivative
evaluated at $\mu_{ex} =0$ yields the canonical fluctuations
$<\delta^{2}N_{0}>_{cn}$.

To proceed, one determines first $\ln{Z_{ex}}$ by the method of
the most probable value. This is done expressing $Z_{ex}$
in terms of the associated generating function, i.e. 
the grand--canonical partition function $\Xi_{ex}$. 
As explained below, this introduces a further 
chemical potential $\mu_{s}$, not to be confused with the 
chemical potential $\mu_{ex}$ of the excited atoms 
introduced in Eq. (1).
The saddle point value $\bar{\mu}_{s}$
of this chemical potential is
fixed by the value $\bar{N}_{ex}$ of $N_{ex}$ for which
$\ln{Z_{ex}}$ acquires its maximum.
Standard calculations
lead to the following expression for the most probable value:
\begin{equation}
\ln{Z_{ex}}(N, \bar{N}_{ex}, \beta)
= -\beta E(N, \bar{N}_{ex}) + \bar{\mu}_{s}\bar{N}_{ex}
- V\left( \frac{m}{2\pi \beta \hbar^{2}} \right)^{3/2}
g_{5/2}(\exp{-\bar{\mu}_{s}}) \; ,
\end{equation}

\noindent where $g_{5/2}$ is the Bose function of order
$5/2$ and 
$\bar{\mu}_{s} = 2\beta \alpha (N-\bar{N}_{ex})$ 
is positive defined. To simplify notations, we have
absorbed a factor $\beta$ in the definition, so
that $\bar{\mu}_{s}$ is adimensional, 
in contrast to the chemical potential $\mu_{ex}$.

Saddle point evaluation of the total free energy 
at the equilibrium value $\bar{N}_{ex}$ can now 
be performed to yield
\begin{equation}
\ln Z_{aux} (N, \beta, \mu_{ex}) =
\ln{Z_{ex}}(N, \bar{N}_{ex}, \beta)
+ \beta \mu_{ex}\bar{N}_{ex} 
+ \frac{1}{2} (\beta \mu_{ex})^{2}
\frac{\partial^{2} \ln{Z_{ex}(N, \bar{N}_{ex}, 
\beta)}}{\partial \bar{N}_{ex}^{2}} \; .
\end{equation}

We see by immediate inspection that the first derivative with 
respect to $\mu_{ex}$ evaluated at the equilibrium value 
$\mu_{ex} =0$ yields the canonical mean value $<N_{ex}>_{cn} 
= \bar{N}_{ex}$, while the second derivative (again
evaluated at $\mu_{ex}=0$) yields the canonical 
root--mean--square fluctuations $<\delta^{2}N_{ex}>_{cn} = 
\partial^{2}\ln{Z_{ex}}/\partial \bar{N}_{ex}^{2}$.

The implicit equation satisfied by the canonical
mean number of condensed atoms $<N_{0}>_{cn} 
\equiv \bar{N}_{0}$ is
$\bar{N}_{ex} = N - \bar{N}_{0}$, where the latter
term reads:
\begin{equation}
N - \bar{N}_{0} = V \left(\frac{m}{2 \pi \beta 
\hbar^{2}}\right)^{3/2} g_{3/2}(\exp(-2\beta \alpha 
(N - \bar{N}_{ex} ))) \, , 
\end{equation}

\noindent where the 
Bose--Einstein function $g_n(s)=\sum_{l=1}^{\infty} s^l/l^n$.
In the same approximation the canonical fluctuations read
\begin{equation}
   <\delta^{2}N_{0}>_{cn} = \left[ \left( \sum_{{\bf p} \neq 0} 
   \frac{1}{4 \sinh^2 
   \left[ \frac{\beta}{2} \left(\frac{{\bf p}^2}{2m} + 
2\alpha (N - \bar{N}_{ex}) \right) \right] } \right)^{-1} -
2 \alpha \beta \right]^{-1} \, .
\end{equation}

\noindent These expressions, as it should be, coincide with
the ones already obtained by Idziaszek {\it et al.} \cite{Kazik}
in the standard canonical setting. Now, by applying 
twice the Legendre transformation we can
express the microcanonical thermodynamic derivatives (fundamental
variables: ${E, N}$) in terms of the canonical ones (fundamental
variables: ${\beta, N}$). One has:
\begin{equation}
<\delta^{2}N_{0}>_{mc} \, \, = \, 
\left. \frac{\partial^{2} 
\ln Z_{aux}}{\partial \mu_{ex}^{2}}\right|_{\mu_{ex}=0;\beta} 
\; \, + \, \, \left. \frac{\partial \beta}{\partial 
\mu_{ex}}\right|_{\mu_{ex}=0;E} \; \, \times \; \left. 
\frac{\partial^{2}\ln Z_{aux}}{\partial \beta 
\partial \mu_{ex}}\right|_{\mu_{ex}=0;\beta} 
\, \, \, .
\end{equation}

The above expression for the microcanonical fluctuations can
be recast by simple thermodynamic manipulations in a
more transparent form. By noting that 
\begin{equation}
\left. \frac{\partial \beta}{\partial 
\mu_{ex}}\right|_{\mu_{ex}=0;E} \; = \, \,
- \left. \frac{\partial E}{\partial 
\mu_{ex}}\right|_{\mu_{ex}=0;\beta} \; \, \times
\; \left. \frac{\partial \beta }{\partial E} 
\right|_{\mu_{ex}=0} \, \, \, ,
\end{equation}

\noindent we can write
\begin{equation}
<\delta^{2}N_{0}>_{mc} \; \; = \, \, <\delta^{2}N_{0}>_{cn}  
\, - \; \frac{ \left[ <\delta N_{0} \delta E>_{cn} 
\right] ^{2}}{<\delta^{2} E>_{cn}} \; .
\end{equation}

Relation (11) is our first main result.
It coincides with the formula derived by Navez {\it et al.}
\cite{Nav} for the ideal Bose gas.  
We thus find that the general structure of
thermodynamics and thermodynamical fluctuations is not
modified by the presence of a model interaction that conserves
the total number of particles. In other words, our analysis
satisfies the general principle that the thermodynamic
relations between different statistical ensembles should
be independent of the dynamics (the Hamiltonian), 
a principle that is violated when resorting to Bogoliubov
theory.

\newpage

Direct analytic evaluation of the microcanonical fluctuations 
is now possible by computing the derivatives 
of the approximate free energy (Eq. (6)) 
appearing in Eq. (9). 
The final result is:
\begin{equation}
<\delta^{2}N_{0}>_{mc} \; = \; <\delta^{2}N_{0}>_{cn}
\; - \;  \frac{\left[A(\beta)F(\beta, 
\bar{\mu}_{s})\right]^{2}}{G(\beta, \bar{\mu}_{s})} \, ,
\end{equation}

\noindent where
\begin{eqnarray}
A(\beta ) & = & V\left(\frac{m}{2\pi \beta 
\hbar^{2}}\right)^{3/2} \, , 
\nonumber \\
& & \nonumber \\
& & \nonumber \\
F(\beta, \bar{\mu}_{s}) 
& = & \frac{3g_{3/2}(\exp{(-\bar{\mu}_{s})}) +
4\alpha \beta \bar{N}_{0}g_{1/2}
(\exp{(-\bar{\mu}_{s})})}{2 \beta [1-2A(\beta ) 
\alpha \beta g_{1/2}(\exp{(-\bar{\mu}_{s})})]} \; , 
\nonumber \\
& & \nonumber \\
& & \nonumber \\
G(\beta, \bar{\mu}_{s}) & = & \frac{3}{2}A(\beta ) 
\left[ \frac{\bar{\mu}_{s}}{\beta^{2}} - 
2 \alpha \frac{\partial \bar{N}_{0}}{\partial 
\beta} \right] g_{3/2}(\exp{(-\bar{\mu}_{s})}) + 
\frac{\bar{\mu}_{s}}{\beta}
\frac{\partial \bar{N}_{0}}{\partial \beta} + 
\frac{15}{4\beta^{2}}A(\beta )
g_{5/2}(\exp{(-\bar{\mu}_{s})}) \, .
\end{eqnarray}

The above expression is our second main result.
It generalizes the microcanonical
fluctuations of the number of
condensed atoms in the ideal Bose gas to the interacting
case. For small values of the coupling $\alpha$
(i.e. of the scattering
length $a$) we can exploit the Robinson--Kac
representation of the Bose $g$ functions \cite{Ziff} 
up to terms linear in $(-\bar{\mu_{s}})^{1/2}$. 
In this situation,
the canonical fluctuations behave normally (scaling linearly
with $V$) as already shown by Idziaszek {\it et al.} \cite{Kazik},
while the second term in the right--hand--side 
of Eq. (12), up to
order ${\bar{\mu}_{s}}^{1/2}$ reduces to
\begin{equation}
- \frac{3A(\beta)}{5g_{5/2}(1)}\left[ g_{3/2}(1)
+ 2\sqrt{\pi}(\bar{\mu}_{s})^{1/2} \right] ^{2} \, .
\end{equation}

\noindent It is straightforward to verify that this
term too scales linearly with $V$, also in the thermodynamic 
limit $V \rightarrow \infty$, $N \rightarrow \infty$, 
$N/V = const$. Therefore the total microcanonical 
fluctuations scale linearly with the volume
of the box, and show no anomalous behavior for
any finite value of the coupling. 
Obviously, in our model the thermodynamic limit
does not commute with the weak coupling limit, and 
to recover the anomalous scaling of the fluctuations
in the ideal Bose gas one must first 
perform the thermodynamic limit and then
go to the limit $a \rightarrow 0$ of vanishing scattering
length. 

In Fig. 1 the microcanonical root--mean--square
fluctuations Eq. (14) are plotted (dotted line)
and compared to the canonical results Eq. (14)
of Ref. \cite{Kazik} (dot--dashed line),
to the canonical fluctuations computed
in the framework of Bogoliubov theory, Eq. (8)
of Ref. \cite{Gior} (dashed line) and to the
ideal Bose gas behavior (solid line). The
fluctuations are plotted as a function of 
the temperature for $N=10^{4}$ particles
and $a/L=5\times 10^{-4}$. The temperature
is measured in units given by the spacing
between the two lowest levels in the 3D box:
$\Delta = (2\pi\hbar)^{2}/(2mL^{2})$.
The inset displays results for $N=10^{6}$
and $a/L=5\times 10^{-5}$; the curves
correspond to our microcanonical
result (dashed), the ideal Bose gas (solid), 
and the canonical results of 
Ref. \cite{Gior} (dashed). The
canonical results of Ref. \cite{Kazik}
are not shown in the inset
because with increasing
number of particles they become
indistinguishable from the microcanonical
prediction, as expected.

We see that our result (14) is
in marked disagreement with the predictions
obtained in \cite{Gior}. The emerging
picture of ideal versus weakly interacting 
Bose systems is then the following.
In the framework of Bogoliubov approximation:
the anomalous scaling behavior of the ideal
gas fluctuations is preserved; the general
thermodynamic relation between ideal gas 
fluctuations computed in different ensembles is lost.
In the framework of number conserving approximations:
the scaling behavior of the ideal Bose gas fluctuations
is modified; the ideal Bose gas thermodynamic
relations are preserved.

To summarize, we have performed a model calculation of
the microcanonical fluctuations of the weakly interacting
Bose--Einstein condensate, allowing for analytical
solution, to provide a first
prediction that may be tested against experiment. 
We stress this point,
since in the real experimental situations the trapped gases
are usually in conditions of fantastic isolation, and thus the
microcanonical result should be the one to be compared
with actual experimental observations (when they will become
available).

At variance with the canonical fluctuations predicted in 
the framework of Bogoliubov theory \cite{Gior}, 
we find no anomalous
scaling of the microcanonical fluctuations
for any finite value of the scattering length.
On the other hand, exact knowledge of the full
energy spectrum
is available only for the one--dimensional chain of
interacting bosons \cite{Lieb}. 
In this case, comparison of the
exact solution with Bogoliubov theory shows that the
Bogoliubov spectrum accounts only for some subset of
the low--lying elementary excitations. The situation 
for Bogoliubov theory is likely to be even worse in 
higher dimensions. Furthermore, phonons are collective
excitations which do not necessarily
change the number of particles
in the condensate. It is therefore hardly conceivable
that they should play an important role 
in determining the equilibrium fluctuations
in the occupation number, and may therefore be safely
ignored in first approximation.

Some of the most subtle aspects of the dynamics
and statistical mechanics of many--body 
interacting Bose systems appear in the above 
discussion: care is needed in handling the 
different limits, whose order cannot in general
be interchanged, while the non--trivial
thermodynamic identities, relating the different
statistical ensembles at the level
of fluctuations, are found to hold true
just as in the case
of the ideal Bose gas \cite{Ziff}.

Looking at the future perspectives, we expect
that achieving a deeper understanding of the
thermodynamic properties in exactly
solvable instances such as the  
Lieb model in one dimension would be of great help,
and studies addressing this problem are currently
under way.
On the experimental side, we hope that 
the measurement of the second--order correlation 
function $g_{2}(r_{1},r_{2})$ will soon  
become available, allowing for a direct test
of the different theoretical predictions.

FI gratefully aknowledges support from
the Alexander von Humboldt Stiftung (Bonn)
and the Istituto Nazionale di Fisica della Materia
(Genua). PN was supported 
by the TMR Network ``Coherent Matter Wave Interactions''
contract ERBFMRX CT96--0002 and by the Republic of Greece
State Scholarship Foundation, contract No. 384. 
MW gratefully aknowledges support by the 
Deutsche Forschungsgemeinschaft.

\vspace{0.4cm}

FIG. 1. Root--mean--square fluctuations of the ground
state occupation number
for $N=10^{4}$ and $a/L=5\times 10^{-4}$.
Displayed are (i) our microcanonical result 
Eq. (14) (dotted), (ii) Eq. (14)
of Ref. \cite{Kazik} (dot--dashed), 
(iii) the ideal Bose gas (solid line),
and (iv) Eq. (8) of Ref. \cite{Gior} (dashed).
Inset: root--mean--square fluctuations
for $N=10^{6}$ and $a/L=5\times 10^{-5}$.
Displayed are (i) our microcanonical result
(dotted), (ii) the ideal Bose gas (solid line),
and (iii) Eq. (8) of Ref. \cite{Gior} (dashed).

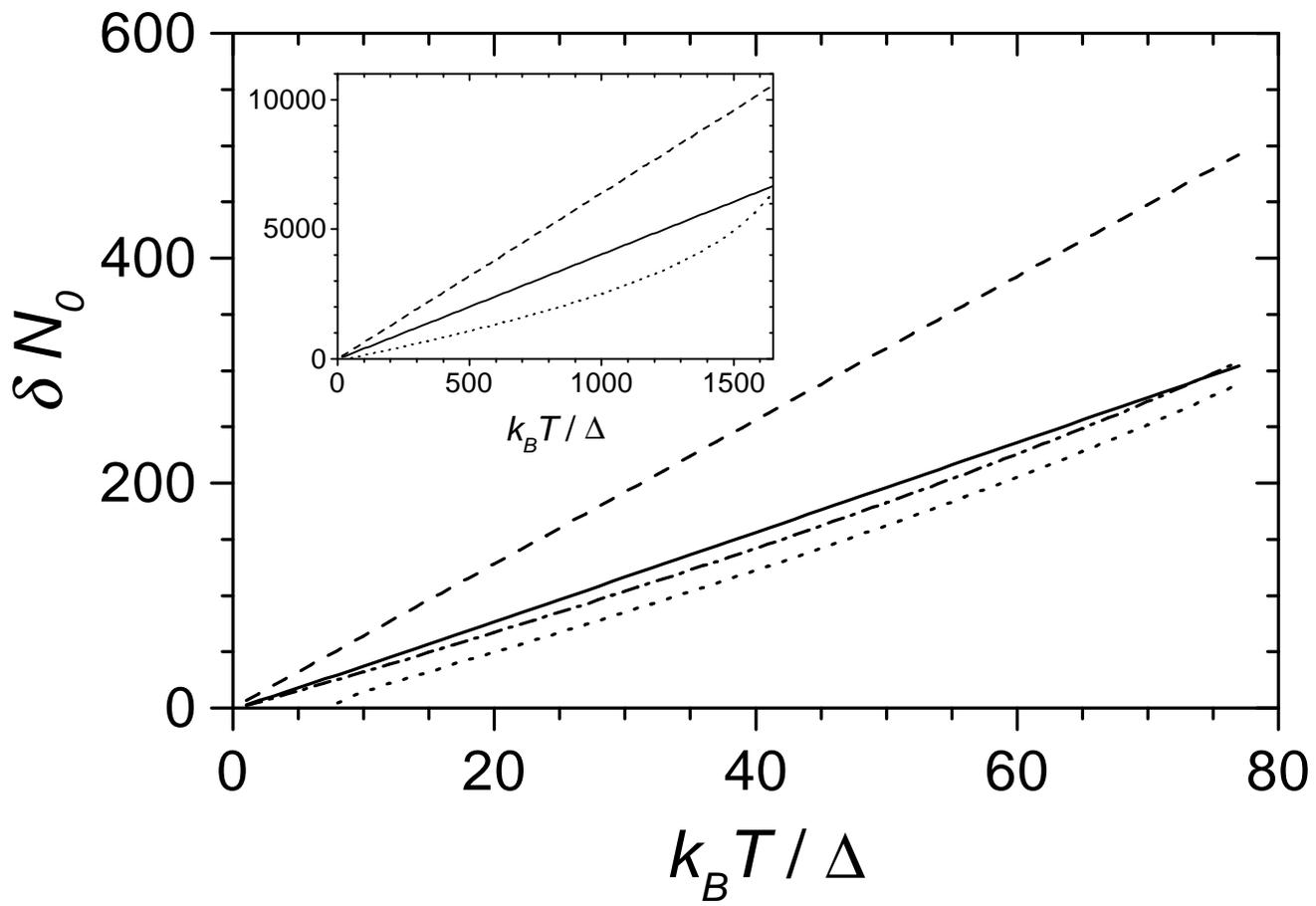
\end{document}